\documentclass[11pt,a4paper]{article}

\usepackage{amsmath,amssymb,wasysym,cite,epsfig,graphicx}

\begin{document}
	
\title{Influence of an anisotropic matter field on the shadow of a rotating black hole}
\author{Javier Bad\'ia$^{1, 2}$\thanks{e-mail: jbadia@iafe.uba.ar}, Ernesto F. Eiroa$^{1}$\thanks{e-mail: eiroa@iafe.uba.ar}\\
	{\small $^1$ Instituto de Astronom\'{\i}a y F\'{\i}sica del Espacio (IAFE, CONICET-UBA),}\\
	{\small Casilla de Correo 67, Sucursal 28, 1428, Buenos Aires, Argentina}\\
	{\small $^2$ Departamento de F\'{\i}sica, Facultad de Ciencias Exactas y Naturales,} \\ 
	{\small Universidad de Buenos Aires, Ciudad Universitaria Pabell\'on I, 1428, Buenos Aires, Argentina}}
\date{}

\maketitle

\begin{abstract}
In this paper, we study the shadow produced by rotating black holes with charge in the presence of an anisotropic matter field. We obtain the apparent shape and the corresponding observables characterizing the size, oblateness, and deviation of the center, for different values of the parameters of the model. We compare the new results with those corresponding to the Kerr-Newmann spacetime and we also discuss the observational prospects.
\end{abstract}

\section{Introduction}

Astrophysical research in the past two decades has accumulated strong evidence for the existence of supermassive black holes at the center of most galaxies, including the Milky Way \cite{gillessen17} and the nearby M87 \cite{broderick15}. One year ago, the international collaboration Event Horizon Telescope (EHT) presented the first reconstructed image \cite{eht19} of the close environment of the supermassive black hole located at the center of the giant elliptical galaxy M87,  obtained by a global very long baseline interferometry (VLBI) array in millimeter radio waves (230 GHz). The image shows the presence of the so-called shadow, surrounded by the light coming from the accretion disk around the black hole M87*, with the bright emission ring having a diameter of $42 \pm 3$ $\mu $as. 
The behavior of the photons in the neighborhood of a black hole results in the shadow or apparent shape as seen by a far away observer. The shadows of non-rotating black holes are circles, but rotating ones show a deformation produced by the spin \cite{bardeen,chandra}. Many researchers have studied this topic in the years previous to the EHT announcement, both in Einstein theory \cite{luminet,falcke00,devries,takahashi,hioki09,bambi09,shadowplas,tsupko17,other18,ghosh20} and in modified gravity \cite{hioki08,amarilla,braneworld,other14,tsukamoto14,perlick14,herdeiro,other16,tsukamoto18,shadowbwcc,ovgun18}. In the case of modified gravity, the size and the shape of the shadow, which always depend on the mass, the angular momentum, and the inclination angle of the black hole, are also related with other parameters specific of the particular model adopted. The EHT discovery has led to an outburst in the number of works published in the field; see, for example, \cite{new19rg,new19at,new20rg,new20at}. It is expected that more detailed direct observations of M87* and also of other black holes will be possible in the coming years \cite{zhu19,millimetron,ehi,observ}, so that the analysis of the shadows will be a useful tool for a better knowledge of astrophysical black holes and also for comparing alternative theories with General Relativity. Other interesting topics concerning the physical nature of the black hole photon ring and the shadow have been recently discussed in the literature \cite{topics}.

The presence of fields or fluids modifies the geometry with respect to Kerr vacuum spacetime, leaving its imprint on the apparent shape of a black hole as seen by a distant observer. Isotropic fluids have been investigated extensively in General Relativity, but anisotropic ones have not drawn much attention until recently. Many works related to anisotropic matter can be found in the literature in different contexts, for example, compact stars, relativistic stellar objects, self-gravitating systems, stellar objects consistent with quark stars, and black holes, among others. The covariant Tolman-Oppenheimer-Volkoff equations for an anisotropic fluid were also recently found. A brief review on this topic with many references is done in \cite{cho18}, where spherically symmetric black holes with a simple anisotropic fluid are introduced. In this article, we investigate the shadow cast by charged rotating black holes under the influence of an anisotropic matter field, assuming a recently obtained solution \cite{kim20}. In Sec. \ref{sec:metric}, the metric of the black hole is reviewed, and the null geodesics are analyzed. The shape of the shadow is obtained in Sec. \ref{sec:shadow}, for different values of the parameters, and the corresponding observables are defined and calculated. Finally, the differences with the Kerr-Newman geometry and the future observational prospects are discussed in Sec. \ref{sec:dis}. We work in units where $G = c = 1$, with $G$ the gravitational constant and $c$ the speed of light.

\section{The black hole metric}\label{sec:metric}

We consider the recently introduced geometry \cite{kim20}, corresponding to a rotating solution of the field equations that result from the action
\begin{equation}
I=  \int  d^4 x \sqrt{-g} \left[\frac{1}{16\pi}(R-F_{\mu\nu}F^{\mu\nu}) +{\cal L}_m \right] ,  \label{action}
\end{equation}
where $R$ is the Ricci scalar, $F_{\mu\nu}$ is the Maxwell electromagnetic field tensor, and ${\cal L}_m$ describes effective anisotropic matter fields, which can be the result of an extra $U(1)$ field as well as diverse dark matter models. The locally anisotropic fluid corresponds to an effective description of matter having a radial pressure not equal to the angular pressure, within the context of General Relativity. The metric, obtained by applying the Newman-Janis algorithm to a static spherically symmetric solution \cite{kiselev03,cho18}, in Boyer-Lindquist coordinates reads \cite{kim20} 
\begin{equation}
	ds^2=  - \frac{\rho^2 \Delta}{\Sigma} dt^2 + \frac{\Sigma \sin^2 \theta}{\rho^2} (d\phi - \Omega\, dt)^2 + \frac{\rho^2}{\Delta} dr^2 + \rho^2 d\theta^2,
\end{equation}
where
\begin{gather}
	\rho^2 = r^2 + a^2 \cos^2 \theta, \\
	\Delta = \rho^2 F(r,\theta) + a^2\sin^2\theta, \label{eq:delta} \\
	\Sigma = (r^2 + a^2)^2 - a^2 \Delta \sin^2\theta, \\
	\Omega = \frac{[1-F(r,\theta)]\rho^2 a}{\Sigma},
\end{gather}
with
\begin{equation}
	F(r, \theta) = 1 - \frac{2Mr - Q^2 + K r^{2(1-w)}}{\rho^2}.
\end{equation}
Here $M$ is the mass, $a = J/M$ the angular momentum per unit mass, and $Q$ the electric charge of the black hole, and the parameters $K$ and $w$, respectively, control the density and anisotropy of the fluid surrounding the black hole \cite{kiselev03,cho18,kim20}. A related rotating black hole spacetime \cite{toshmatov17} was obtained by applying the Newman-Janis algorithm to the solution \cite{kiselev03}, with quintessential energy but without the electromagnetic field (no charge).   When $K = 0$ in the equations above, we recover the Kerr-Newman metric. If the charge term comes from the presence of an electromagnetic field,  then $Q^2$ is clearly positive. However, it is easy to also allow for negative values of $Q^2$; that is, replace $Q^2$ by a parameter $q$ which may be either positive or negative. Such a metric  which in the $K=0$ case may be dubbed ``Kerr-Newman-like'', can arise from alternative theories of gravity or from different kinds of fields surrounding the black hole, with the ``charge'' term not coming from the electromagnetic Lagrangian; see, for example, \cite{braneworld} and the references therein. We will therefore rewrite the function $F(r, \theta)$ in the form
\begin{equation}
	F(r, \theta) = 1 - \frac{2Mr - q + K r^{2(1-w)}}{\rho^2} \label{eq:F},
\end{equation}
where $q$ may in principle take any real value, subject only to the condition that there exists an event horizon, in order that the geometry corresponds to a black hole.

In an orthonormal frame $(e_{\hat{t}}, e_{\hat{r}}, e_{\hat{\theta}}, e_{\hat{\phi}})$, in which the stress-energy tensor of the anisotropic matter field is diagonal $T_{\hat{\mu}\hat{\nu}}= \mathrm{diag}(\varepsilon,p_{\hat{r}},p_{\hat{\theta}},p_{\hat{\phi}})$, the expressions for the energy density $\varepsilon$, the radial pressure $p_{\hat{r}}$, and the angular pressures $p_{\hat{\theta}}$ and $p_{\hat{\phi}}$ are \cite{kim20}
\begin{equation}
\varepsilon=\frac{q + (1-2w)K r^{2(1-w)}}{8\pi \rho^{4}}, 
\end{equation}
\begin{equation}
p_{\hat{r}}=-\varepsilon ,
\end{equation} 
\begin{equation}
p_{\hat{\theta}}=p_{\hat{\phi}}
	= \left( \rho^{2}w-a^2\cos^2\theta\right) \frac{\varepsilon}{r^2} + (1-w)\frac{q}{8\pi \rho^2 r^2}.
\end{equation}
There are some restrictions one can impose on the parameters $w$ and $K$. The metric is clearly only asymptotically flat for $w>0$, and we consequently restrict ourselves to that case. In addition, the total energy is finite only when $w > 1/2$, because otherwise the energy density is not localized sufficiently, and the condition $q + (1-2w)K r^{2(1-w)} \geq 0$ must hold to have a non-negative energy density at a radius $r$ in the rest frame of the matter surrounding the black hole \cite{kim20}. This, however, does not affect the calculation of the black hole shadow, and hence we will only demand that $w$ is positive, allowing $K$ to take either sign. 

\subsection{Event horizon}\label{sec:eh}

The event horizon of this spacetime is located at the largest radius for which $\Delta(r) = 0$. After substituting the expression \eqref{eq:F} for $F(r,\theta)$ in the definition \eqref{eq:delta} of $\Delta$, we find
\begin{equation}
\begin{split}
	\Delta(r) &= r^2 + a^2 + q - 2Mr - K r^{2(1-w)} \\
	&\equiv \Delta_{KN} - K r^{2(1-w)},
\end{split}
\end{equation}
where $\Delta_{KN} = r^2 + a^2 + q -2Mr$ is the functional form of $\Delta$ in a Kerr-Newmann-like spacetime.

\begin{figure}[t!]
	\centering
	\includegraphics[width=\textwidth]{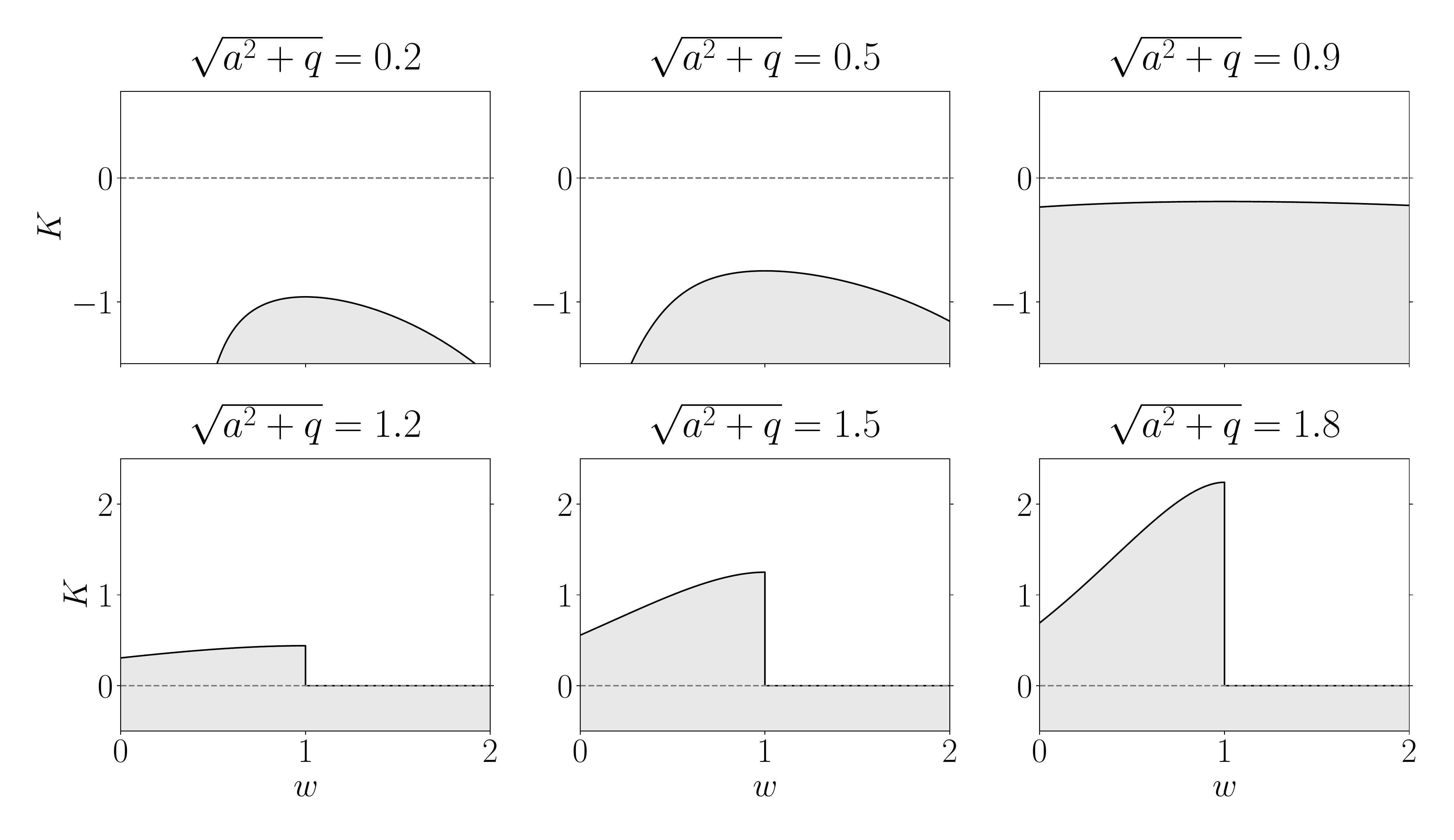}
	\caption{The allowed values of $w$ and $K$. The solid curves show the parameter values for which the event horizon disappears: in the shaded regions below them, the spacetime contains a naked singularity. As mentioned in the text, we take $M=1$, so that $K$ is dimensionless.}
	\label{fig:horizontes}
\end{figure}

We want to find the regions in parameter space for which an event horizon exists. The disappearance of the event horizon corresponds to a double root of $\Delta$, i.e., a simultaneous solution of the system of equations
\begin{equation}
	\begin{cases}
	r^2 + a^2 + q - 2Mr - K r^{2(1-w)} &= 0 \\
	2(r-M) - 2(1-w)K r^{1-2w} &= 0.
	\end{cases}
\end{equation}
Except for special values of $w$ this system cannot be solved for $r$ in closed form, but we can easily find parametric expressions for $K$ and $w$ as functions of $r$, assuming that $M$, $a$, and $q$ have been fixed beforehand. In fact, the solutions depend on $a$ and $q$ only through the combination $a^2+q$, which reduces by one the number of independent parameters. To start, we simply solve for $K$ in the first equation to obtain $K = \Delta_{KN} r^{2(w-1)}$; then, by plugging this expression into the second equation, we find
\begin{equation}\label{eq:wcrit}
	w = \frac{a^2 + q - Mr}{\Delta_{KN}},
\end{equation}
from which it follows immediately that
\begin{equation}\label{eq:kcrit}
	K = \frac{\Delta_{KN}}{r^{2r(r-M)/\Delta_{KN}}}.
\end{equation}
These relations are plotted in Fig. \ref{fig:horizontes}. The curve shown separates the spacetimes with a naked singularity corresponding to the shaded region, from those having the presence of the event horizon, represented by the region above this curve. In fact, the full graph of Eqs. \eqref{eq:wcrit} and \eqref{eq:kcrit} is more complex than shown in the picture, because there are additional interior horizons that can appear or disappear. Since we are not interested in the black hole interior, we have simply shown the separation curve between the naked singularities and the spacetimes with an event horizon. Note that it is possible to have $|a| > 1$, and in fact any value of $a$ at all, if $K$ and $w$ are chosen appropriately.
Some care is required when graphing because the dimensions of $K$ depend on $w$, and therefore values of $K$ for different values of $w$ cannot be directly compared. More concretely, from the definition of $F(r,\theta)$ we see that $K r^{-2w}$ must be dimensionless, so that $K$ has dimensions of length raised to the $2w$ power. In our plots, we take $M=1$ throughout, making all quantities dimensionless; this is equivalent to replacing $K$ by a new dimensionless parameter $\tilde{K} = K / M^{2w}$.

We can see that Eqs. \eqref{eq:wcrit} and \eqref{eq:kcrit} together with Fig. \ref{fig:horizontes} show how to find the values of $w$ and $K$ at which the event horizon disappears, for given values of $M$, $a$, and $q$. In this work, however, we will be plotting various quantities as functions of $q$, and so we are mostly concerned with the opposite problem: solving for the value of $q$ (or, as mentioned before, of $a^2+q$) at which the event horizon disappears, for given values of $K$ and $w$. This amounts to finding the values of $r$ and $q$ at which both $\Delta(r,q)$ and $\partial_r \Delta(r,q)$ equal zero, and should be done numerically.

\subsection{Null geodesics}\label{sec:ngeo}

To find the geodesics in this spacetime we follow the standard method, first introduced in  \cite{carter} for the Kerr geometry (see, for example, \cite{chandra}), of separating variables in the Hamilton-Jacobi equation
\begin{equation}
\frac{\partial S}{\partial \lambda}=-\frac{1}{2}g^{\mu\nu}\frac{\partial S}{\partial x^\mu}\frac{\partial S}{\partial x^\nu},
\end{equation}
where $\lambda$ is an affine parameter along the geodesics and $S$ is the Jacobi action. For the geometry considered here, this action is separable in the simple form
\begin{equation}
S=\frac{1}{2}\mu^2 \lambda -E t + L\phi + S_r(r) + S_{\theta}(\theta),
\end{equation}
with $\mu$ the mass of a test particle, $E$ the energy, and $L$ the azimuthal angular momentum. The quantities $E$ and $L$ are constants of motion along the geodesic, related to the symmetries of the spacetime, with cyclic coordinates $t$ and $\phi$, and the associated Killing vectors. The functions $S_r(r)$ and $S_{\theta}(\theta)$ only depend on $r$ and $\theta$, respectively. Considering null geodesics, i.e., $\mu =0$, the Hamilton-Jacobi equation
can be rewritten in the form
\begin{equation}
-\Delta \left(\frac{d S_r}{dr}\right)^2 + \frac{\left[(r^2+a^2)E-aL \right]^2}{\Delta} = \left(\frac{d S_\theta}{d\theta}\right)^2 + \frac{\left(L-aE\sin^2\theta \right)^2}{\sin^2\theta }.
\end{equation}
In this equation, one side is a function only of $r$ and the other side depends only on $\theta$; therefore, both sides should be equal to the same positive constant, denoted by $\mathcal{K}$. Using that $dx^\mu /d\lambda =p^\mu = g^{\mu \nu } p_\mu$ and $p_\mu = \partial S / \partial x^\mu $, the separation procedure leads in our case to the first-order geodesic equations
\begin{align}
	\rho^2 \dot{t} &= \frac{r^2+a^2}{\Delta} P(r) - a(a \sin^2\theta E - L), \\
	\rho^2 \dot{\phi} &= \frac{a P(r)}{\Delta} - aE + \frac{L}{\sin^2 \theta}, \\
	\rho^2 \dot{r} &= \pm \sqrt{\mathcal{R}(r)}, \\
	\rho^2 \dot{\theta} &= \pm \sqrt{\Theta(\theta)},
\end{align}
where
\begin{align}
	P(r) &= E(r^2+a^2) - aL, \\
	\mathcal{R}(r) &= P(r)^2 - \Delta [(L-aE)^2 + \mathcal{Q}],
\\
	\Theta(\theta) &= \mathcal{Q} + \cos^2\theta \left(a^2 E^2 - \frac{L^2}{\sin^2\theta}\right);
\end{align}
in them, the dot represents the derivative with respect to $\lambda$ and $\mathcal{Q}=\mathcal{K}-(L-aE)^2$ is the Carter constant \cite{carter}. Since null geodesics are unaffected by a rescaling of the affine parameter, we rescale the conserved quantities to obtain the impact parameters
\begin{equation}
	\xi= \frac{L}{E}, \qquad \eta = \frac{\mathcal{Q}}{E^2},
\end{equation}
which are independent of the chosen affine parametrization.

\section{Black hole shadow}\label{sec:shadow}

We now proceed with the study of the apparent shape of the black hole in the sky of a far away observer.

\subsection{Shape of the shadow}\label{sec:shape}

The shadow of a black hole as seen by a distant observer is the set of directions in the sky from which no light rays can arrive from infinity. Its contour is then the border between those trajectories that when propagated backward in time fall into the black hole and those that reach infinity. It corresponds to the trajectories that asymptotically approach the spherical photon orbits of constant $r$ and which therefore have the same impact parameters as them. We review here the procedure, described in detail in \cite{tsukamoto18}, for finding these trajectories for a certain class of metrics. To begin, we note that we can write the function $\Delta$ appearing in the metric as
\begin{equation}
	\Delta = r^2 - 2 m(r) r + a^2,
\end{equation}
with
\begin{equation}
	m(r) = M - \frac{q}{2r} + \frac{K}{2r^{2w-1}}.
\end{equation}
The metric is otherwise identical to the Kerr metric and reduces to it if we set $m(r) \equiv M$. To find the orbits of constant radius, we look for double roots of the radial potential $\mathcal{R}$, which together with its derivative can be expressed as
\begin{align}
	\frac{\mathcal{R}(r)}{E^2} &= r^4 + (a^2 - \xi^2 - \eta)r^2 + 2 [(\xi-a)^2 + \eta] m(r) r - a^2 \eta, \\
	\frac{\mathcal{R}'(r)}{E^2} &= 4 r^3 + 2 (a^2 - \xi^2 - \eta) r + 2 [(\xi-a)^2 + \eta] (m(r) + m'(r)r).
\end{align}
Setting $\mathcal{R}(r) = 0 = \mathcal{R}'(r)$ gives a pair of equations which is quadratic in the impact parameters $\xi$ and $\eta$, and thus can be solved for them as parametric functions of the radius $r$ of the spherical photon orbit. This yields two solutions, one of which is not relevant for the black hole shadow (see \cite{chandra,tsukamoto18} for details); the other solution is
\begin{gather}
\xi = \frac{4 m(r) r^2 - (r + m(r) + m'(r)r)(r^2 + a^2)}{a(r - m(r) - m'(r)r)}, \label{eq:xiC} \\
\eta = r^3 \frac{4a^2 (m(r)-m'(r)r) - r(r - 3m(r) + m'(r)r)^2}{a^2 (r - m(r) - m'(r)r)^2}. \label{eq:etaC}
\end{gather}

\begin{figure}[t!]
	\centering
	\includegraphics[width=\textwidth]{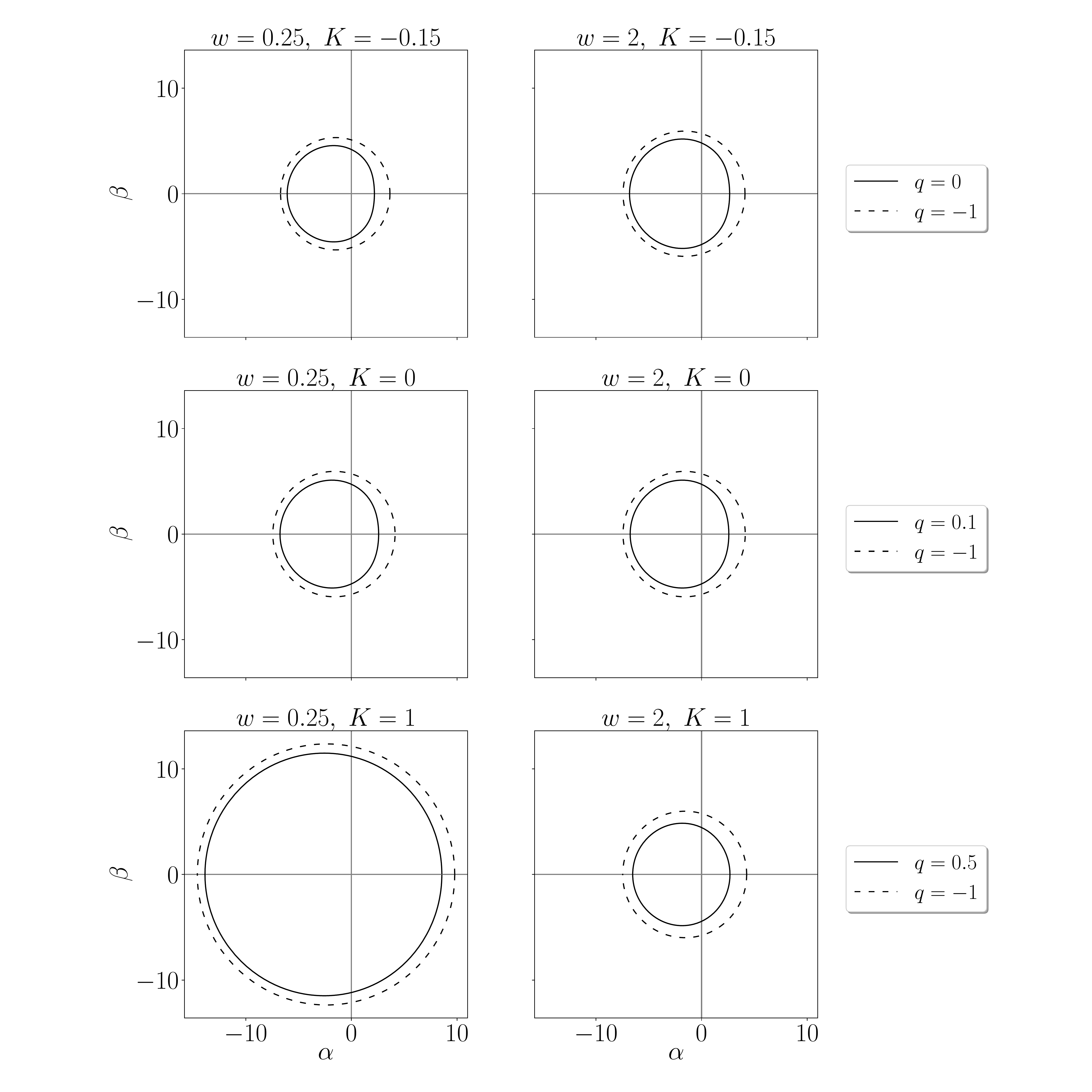}
	\caption{Contour of the black hole shadow for $a = 0.9$ and $\theta_\text{o} = \pi/2$; all quantities have been adimensionalized by setting $M=1$.}
	\label{fig:sombras}
\end{figure}

Assuming an observer at infinity, we adopt the celestial coordinates \cite{bardeen,chandra}
\begin{gather}
	\alpha = - r_0 \frac{p^{\hat{\phi}}}{p^{\hat{t}}} \bigg|_{r_0 \to \infty}, \\
	\beta = - r_0 \frac{p^{\hat{\theta}}}{p^{\hat{t}}} \bigg|_{r_0 \to \infty},
\end{gather}
where $p^{\hat\mu}$ are the components of the momentum in the orthonormal tetrad of the observer; the orientation is such that the $\beta$ axis is aligned with the spin of the black hole. Writing this momentum in terms of the impact parameters, the coordinates of an incoming photon can then be shown to be \cite{bardeen,chandra}
\begin{gather}
	\alpha = - \frac{\xi}{\sin\theta_\text{o}}, \\
	\beta = \pm \sqrt{\eta + \cos^2\theta_\text{o} \left(a^2 - \frac{\xi^2}{\sin^2\theta_\text{o}}\right)},
\end{gather}
where $\theta_\text{o}$ is the inclination angle of the observer from the spin axis. Using the expressions \eqref{eq:xiC} and \eqref{eq:etaC} for the critical impact parameters, these equations give a curve in the $(\alpha, \beta)$ plane parametrized by the radius $r$ of the spherical photon orbit corresponding to each point. The domain is bounded by the radii $r_\pm$ for which $\beta(r_\pm) = 0$, which should be found numerically.

Some example contours are shown in Fig. \ref{fig:sombras}. The usual effects of the rotation of the black hole are clearly seen: for non-null $a$, the shadow is asymmetrical and displaced from the origin of coordinates. To show these effects as clearly as possible, in this and in subsequent figures we have chosen to place the observer at the equatorial plane, with $\theta_\text{o} = \pi/2$. We recall at this point that even though the shadow can be found for any values of the parameters $w$ and $K$, physical considerations impose some restrictions: as mentioned in Sec. \ref{sec:metric}, the total energy in the rest frame of the surrounding matter will only be finite for $w > 1/2$, and the energy density will only be non-negative at radii for which $q + (1-2w)K r^{2(1-w)} \geq 0$. The dependence of the shape of the shadow on the parameters is discussed in the next section.

\subsection{Observables}\label{sec:obs}

\begin{figure}[t!]
	\centering
	\includegraphics[width=\textwidth]{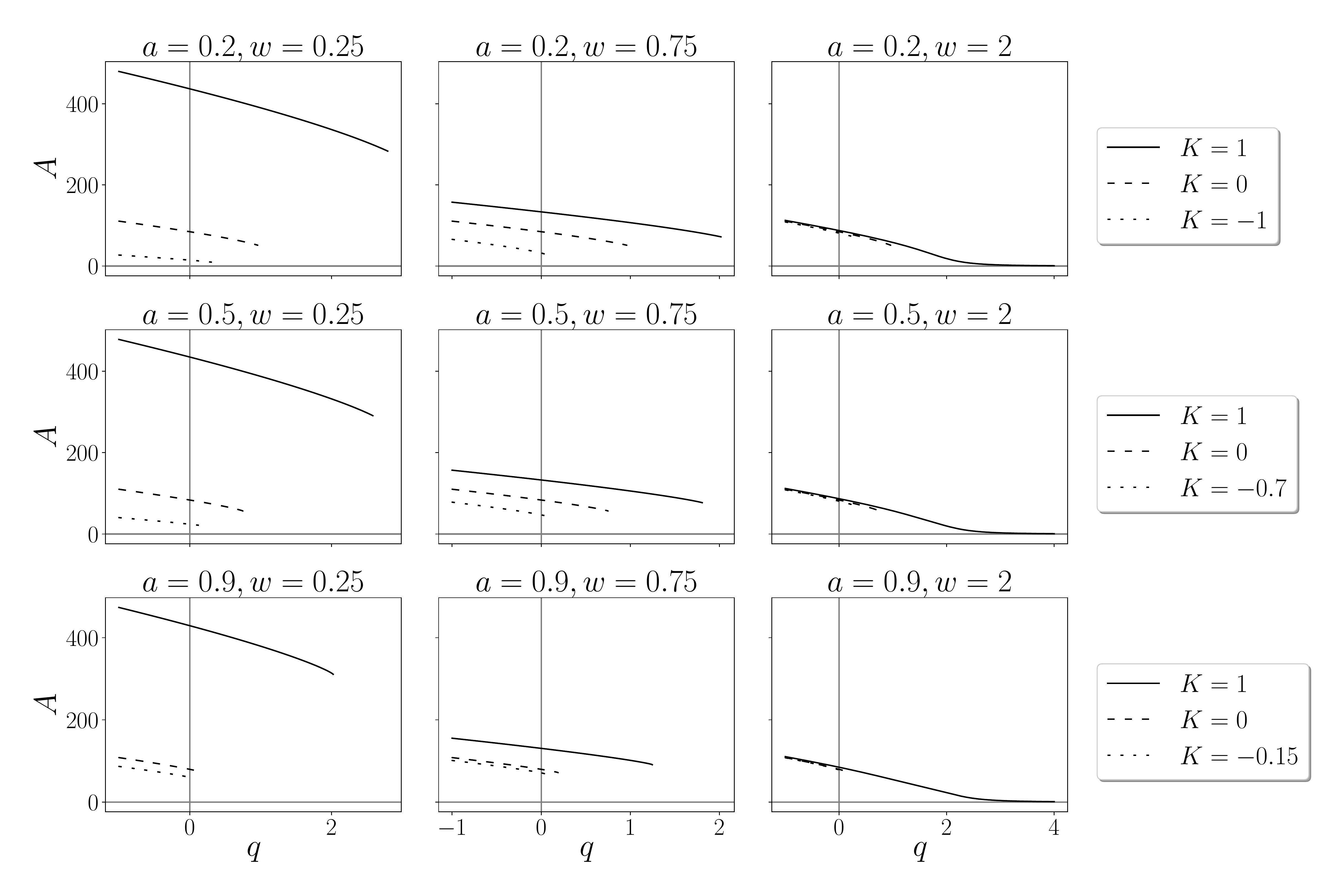}
	\caption{The area of the black hole shadow for some representative values of the parameters. All quantities have been adimensionalized by setting $M=1$.}
	\label{fig:areas}
\end{figure}

\begin{figure}[t!]
	\centering
	\includegraphics[width=\textwidth]{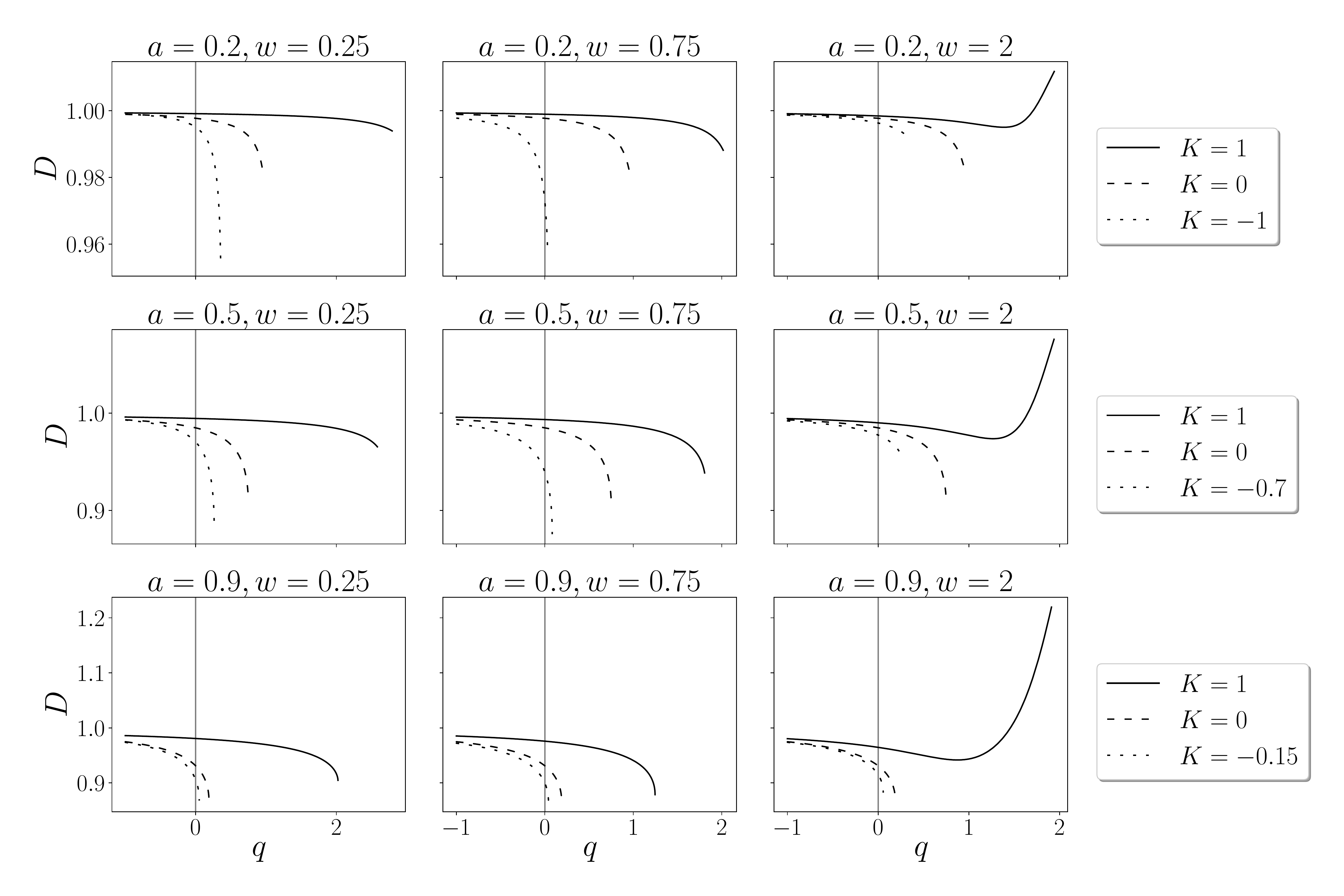}
	\caption{The oblateness of the black hole shadow for some representative values of the parameters.}
	\label{fig:elipt}
\end{figure}

\begin{figure}[t!]
	\centering
	\includegraphics[width=\textwidth]{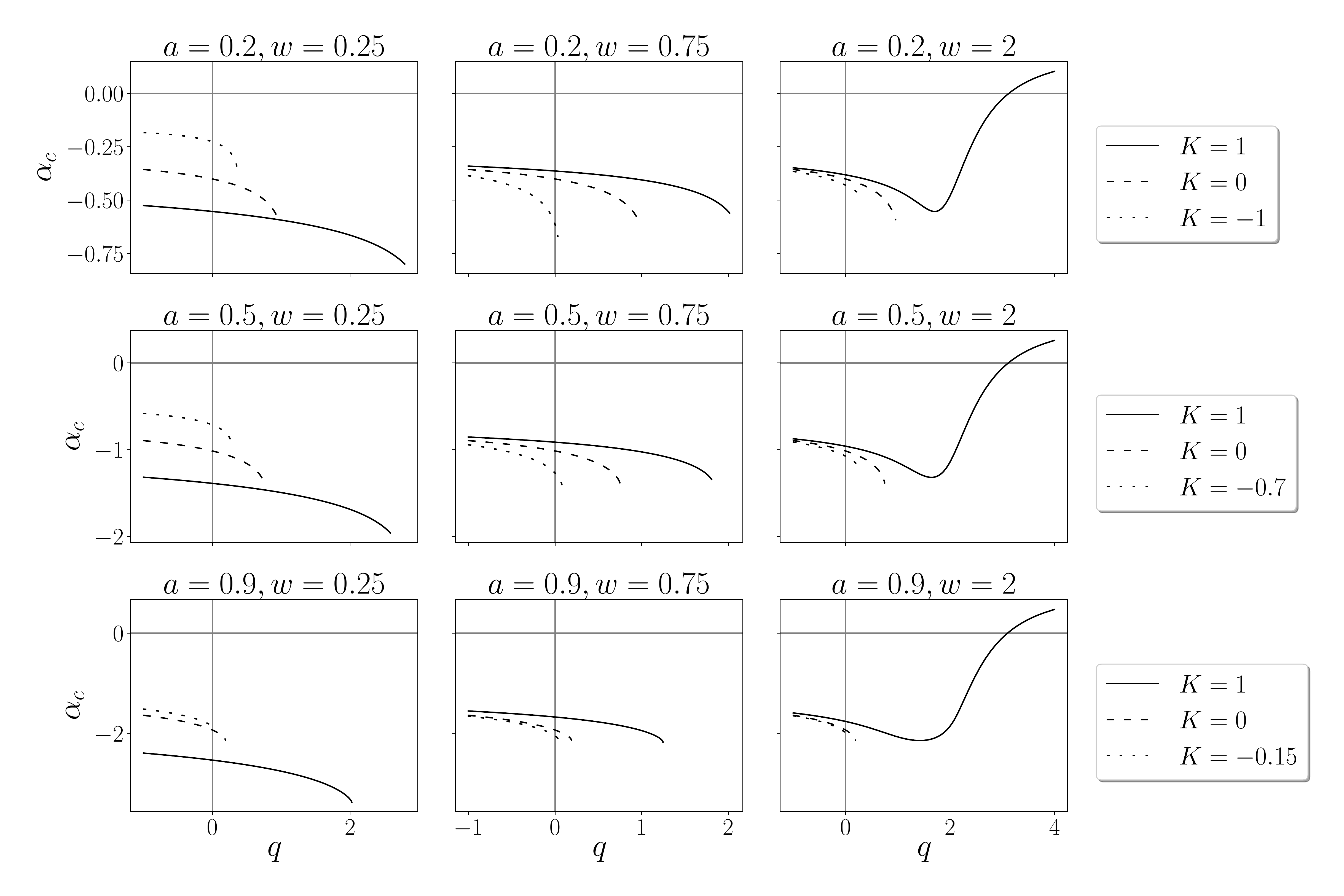}
	\caption{The horizontal displacement of the center of the shadow.}
	\label{fig:despl}
\end{figure}

Various observables have been proposed in the literature \cite{hioki09,tsukamoto14,tsupko17,ghosh20} as a way to describe the shape of the black hole shadow and its dependence on the parameters. We start with the area of the shadow, defined as
\begin{equation}
	A = 2 \int_{\alpha_-}^{\alpha_+} \beta\, d\alpha = 2 \int_{r_+}^{r_-} \beta(r) |\alpha'(r)|\, dr,
\end{equation}
with a factor of two since the curve $(\alpha(r), \beta(r))$, choosing the positive sign for $\beta$, only describes half the shadow. Here $\alpha_-$ and $\alpha_+$ are the coordinates of the left and right ends of the shadow, respectively, and $r_\pm$ the corresponding values of $r$. With standard software, this integral can be calculated numerically, but instead it turned out to be faster to find the area by solving the differential equation $A'(r) = 2 \beta(r) |\alpha'(r)|$.

The dependence of the area on the parameters $a$, $K$, $w$, and $q$ is shown in Fig. \ref{fig:areas}; as mentioned before, we consider negative and positive values of $q$, up to a maximum value for which the event horizons disappear, and we set $M=1$.  It can be seen that the area is in all cases a decreasing function of $q$, generalizing the behavior of the Kerr-Newman black hole; it is also an increasing function of $K$, which is reasonable since $K$ and $q$ appear with opposite signs in the metric. The middle curve, with $K=0$, corresponds to the Kerr-Newman black hole and is independent of $w$. It might be tempting to say that the three curves get closer together as $w$ increases; however, one should remember that $K$ has dimensions of mass raised to the power $2w$, so that values of $K$ for different values of $w$ are not directly comparable.

To quantify the deformation of the shadow, we can also define the oblateness
\begin{equation}
	D = \frac{\Delta \alpha}{\Delta \beta},
\end{equation}
where $\Delta \alpha$ and $\Delta \beta$ are the extent of the shadow in the horizontal and vertical directions respectively; the circular shadow of a non-rotating black hole has $D=1$. In our case, it is plotted as a function of $q$ in Fig. \ref{fig:elipt} for some representative values of the parameters. The dependence on the charge is in this case rather weak and more pronounced for larger spins, with the curves being barely distinguishable for $a = 0.2$, where the shadow is nearly circular. We see that the shadow becomes less circular as $K$ becomes more negative, where the black hole is closer to being extremal, but the difference in oblateness between the lowest and highest values of $K$ stays below $\sim 15\%$. The exception is when $K$ is positive and $w$ is greater than one, which allows $q$ to become arbitrarily large. In this case, we have $D > 1$, and the shadow becomes horizontally stretched instead of compressed.

Last, we consider the displacement between the optical axis and the centroid of the shadow, which lies on the $\alpha$ axis at a position $\alpha_c$ defined by the integral
\begin{equation}
	\alpha_c = \frac{1}{A} \int_{\alpha_-}^{\alpha_+} 2 \alpha \beta\, d\alpha =  \frac{1}{A} \int_{r_+}^{r_-} 2 \alpha(r) \beta(r) |\alpha'(r)|\, dr.
\end{equation}
From an observational perspective, this is the most difficult quantity to measure, since it requires an independent knowledge of the true position of the black hole in the sky. Its dependence on the parameters is shown in Fig. \ref{fig:despl}. Here we see for the first time non-monotonic behavior: for $w = 0.25$ the shadow is displaced further to the left as $K$ increases, while the opposite occurs for $w \geq 0.75$. In fact, the reversal in behavior is gradual, in the sense that it does not happen at a single value of $w$. Rather, as $w$ increases toward $1/2$, the curves begin to cross each other; when $w = 1/2$, all three curves cross at $q=0$: in other words, the displacement for $q=0$ is independent of $K$. As $w$ increases past $1/2$, the crossing points move toward negative values of $q$, as can be seen in the right column of Fig. \ref{fig:despl}. As in the case of the oblateness, we also see drastically different behavior when $q$ becomes large enough: in this case, the shadow eventually moves to the right of the optical axis.

\section{Discussion}\label{sec:dis}

In this work, we have analyzed a spacetime that describes a black hole with electric charge $Q$ surrounded by a perfect fluid with anisotropic pressure.  However, due to its generality, it can have a wide variety of physical origins \cite{kim20}, all describing asymptotically flat black holes as long as the conditions in Sec. \ref{sec:eh} are satisfied. In particular, we have found that for $w>0$, which is the asymptotically flat case, we can have $a > M$ as long as the fluid parameters $K$ and $w$ take values in the allowed region for the presence of the event horizon. This suggests that an observation of a black hole with a spin greater than the Kerr bound could be evidence toward the presence of an additional term in the metric.

Our basic assumption is that the matter does interact with the photons only gravitationally. If the matter absorbs or scatters the photons, our study applies to a frequency band where these effects can be neglected, which depends on the particular details of the matter model. The shadow of a black hole geometry with the same form as the one considered here has been described before in the absence of electric charge \cite{kumar20}, albeit with a different physical motivation for the $K$-dependent term. We have extended the analysis to include the possibility of an electrically charged black hole, or, as mentioned in Sec. \ref{sec:metric}, other Kerr-Newman-like spacetimes, characterized by a parameter $q$ having any sign, instead of $Q^2$. We are interested in the possibility of finding non-Kerr features in a hypothetical observation of a black hole shadow, and to this end we have calculated the values of three observables, assuming an equatorial observer to provide some numbers: the area of the shadow, its oblateness or departure from circularity, and the displacement of the center of the shadow from the optical axis. The latter is probably the most difficult to determine observationally, since it requires an independent measurement of the true position of the black hole in the sky. Still, all three observables can be determined from an observation of a black hole shadow, and they can be used to contrast alternatives to the Kerr metric.

We have found that the properties of the shadow mostly mirror the Kerr-Newman case: as the charge increases, the shadow becomes smaller, the oblateness becomes less than unity, and the shadow moves to the left of the optical axis. This is observed independently of the parameters $w$ and $K$. Increasing $K$ gives the opposite effect to increasing the charge, which is reasonable since the charge term and the $K$ term in the metric have similar behavior, and they appear with opposite signs. As it has been noted, this is however not the case for the displacement of the center of the shadow in the $w < 1/2$ region: increasing $K$ moves the center to the left, as does increasing $q$. There is another exception in the $K>0$, $w>1$ region, where $q$ does not have a maximum value: in this case, the oblateness and the horizontal displacement reverse their behavior and start increasing for large values of $q$. Geometrically, the shadow moves to the right of the optical axis, and its horizontal diameter becomes larger than its vertical diameter. This regime is particularly interesting because it is easily distinguishable from the Kerr shadow and is thus a prime candidate for comparison with future observations by the EHT or other forthcoming instruments.

In the examples presented throughout this work, we have only considered the case of an equatorial observer, but it is straightforward to extend our results to observers with other inclination angles, for which similar characteristic features of the shadow are obtained. Also, for theoretical purposes and completeness, we have taken values of the parameters $q$, $K$, and $\omega$ in a range much larger than physically expected in nature.  However, the reasonably small deviations from the Kerr case expected in our work would not be observable with the current or upgraded EHT instruments and will surely demand more advanced facilities.

Future observations by the EHT and subsequent analysis will allow to study the stability, shape and depth of the shadow with greater precision. One of its key characteristics is that it must remain constant over time, since the mass of M87* is not expected to change on human time scales. Polarimetric image analysis will provide information on the accretion rate and magnetic flux. The black hole Sgr A* has a mass three orders of magnitude smaller than that of M87* and dynamic time scales of minutes instead of days. The observation of the shadow of Sgr A* will require taking into account this variability and mitigating the dispersion effects caused by the interstellar medium \cite{zhu19}. Higher resolution images can be achieved by going to a shorter wavelength, for example, to $0.8$ mm ($345$ GHz), by adding telescopes and, in a farther future, with interferometry with space instruments. The distances between the telescopes in space are larger and they can operate at higher radio frequencies --filtered out by the atmosphere for instruments on Earth-- resulting in a better resolution of images. Millimetron \cite{millimetron} is a planned space-based mission that will operate from far infrared to millimeter bands (in principle, up to 900 GHz), with an expected resolution as a ground-space system of $0.1$ $\mu$as or better. Another recently proposed space VLBI mission \cite{ehi} is the Event Horizon Imager (EHI), that will work at high frequencies (up to $690$ GHz), allowing for extremely high-resolution and high-fidelity imaging of radio sources, and may be suitable for Sgr A*. Proposed x-ray instruments would also have an improved resolution that will allow a detailed exploration of galactic centers in a more distant future.  Other interesting observational aspects are discussed in \cite{observ}. The comparison between the observed shadow of black holes and the theoretical models will be a valuable tool in forthcoming astrophysics.

\section*{Acknowledgments}

This work has been supported by CONICET and Universidad de Buenos Aires.

\end{document}